\documentclass[pra,twocolumn,showpacs,preprintnumbers,amsmath,amssymb,floatfix,superscriptaddress] { 
revtex4-1}

\usepackage{graphicx}
\usepackage{dcolumn}
\usepackage{bm,bbm}

\begin{document}


\title{Propagation of ultra-short, resonant, ionizing laser pulses 
in rubidium vapor}%

\author{Gabor Demeter}
\email{demeter.gabor@wigner.mta.hu}
\affiliation{Wigner Research Center for Physics, Hungarian Academy of Sciences, Konkoly-Thege
Mikl\'os \'ut 29-33, H-1121 Budapest, Hungary}

\date{\today}

\begin{abstract}
We investigate the propagation of ultra-short laser pulses in atomic rubidium vapor. 
The pulses are 
intensive enough to ionize the atoms and are directly resonant with the 780 nm $D_2$ line. We 
derive a relatively simple theory for computing the nonlinear optical response of atoms and 
investigate the competing effects of strong resonant nonlinearity and ionization in the medium 
using computer simulations. A nonlinear 
self-channeling of pulse energy is found to produce a continuous plasma channel with complete 
ionization. We evaluate the length, width and homogeneity of the resulting plasma 
channel for various values of pulse energy and initial focusing to identify regimes optimal for 
applications in plasma-wave accelerator devices such as that being built by the AWAKE collaboration 
at 
CERN. Similarities and differences with laser pulse filamentation in atmospheric gases are 
discussed.  

\end{abstract}

\maketitle

\section{Introduction}
\label{intro}

The propagation of femtosecond laser pulses in various optical media is an active field of study 
with many applications. In particular, pulses powerful enough to ionize atoms and molecules of 
gases they propagate through have been studied intensely in the past 2-3 decades. Their propagation 
is governed by the dynamical competition between optical nonlinearities of various orders, 
intensity clamping due to multiphoton ionization and refractive index changes due to plasma 
generation. The  competition between self-focusing and de-focusing effects lead to the 
formation of filaments, i.e. long, extended domains along the pulse propagation direction 
with strong localization in the transverse plane where gas is ionized. The precise mechanisms 
through which these plasma channels are created and light filaments maintained have been 
investigated extensively both theoretically and experimentally 
\cite{Berge1998,Couairon2007,Berge2007,Kandidov2009,Kolesik2013}. 

A very similar problem arose recently in the context of the 
Advanced Proton Driven Plasma Wakefield Acceleration Experiment (AWAKE)
 experiment at CERN. AWAKE is a     
proton-driven plasma wakefield acceleration experiment, the first of its kind, which uses 
high-energy proton bunches to drive wakefields in a plasma for electron acceleration
\cite{Caldwell2016,Gschwendtner2016,Adli2018}. Central to the device is a 10 m long rubidium vapor 
cell where the proton bunch interacts with the plasma serving as an energy exchange medium 
between the protons and the injected electrons. Under appropriate 
conditions, the self-modulation instability breaks up the proton bunch which then resonantly drives 
the plasma wakefields. Important factors for success are high plasma homogeneity as well as a 
quasi-instantaneous plasma creation for seeding the instability during the time the proton bunch is 
in the cell. This is achieved by ionizing the rubidium vapor in the temperature controlled cell 
by a powerful fs 
laser pulse propagating simultaneously with the proton bunch. The problem is at 
first sight almost identical to filamentation studies in atmospheric gases as the formation of a 
long plasma channel is required by a powerful, ultra-short laser pulse.  

But there are also some fundamental differences. First of all, the 780 nm AWAKE laser is directly 
resonant with the $D_2$ line of rubidium, the transition between the $5\mathrm{S}_{1/2}$ ground 
state to the $5\mathrm{P}_{3/2}$ excited state and very close to resonance with the 
776 nm $5\mathrm{P}_{3/2}\rightarrow 5\mathrm{D}_{5/2},5\mathrm{D}_{3/2}$ transitions. This means 
that there is very strong nonlinear optical interaction between the pulse and the vapor at 
arbitrarily low intensities. Because this nonlinearity is much stronger than the 
ones given by the usual 
nonresonant nonlinear optical coefficients, we get a sizeable response from the medium even though 
the initial vapor density is 
$10^{14}-10^{15}\mathrm{~1/cm^3}$, orders of magnitude less dense than atmospheric gases.   
The effect of such a single-photon resonance is completely missing from usual 
filamentation studies, though  the effects of resonant two- and three-photon 
transitions on the process have been investigated recently
\cite{Doussot2016,Doussot2017}.
Second, high plasma homogeneity is required which must be achieved through 100 \% 
ionization of the initially homogeneous vapor - this means that plasma density gradients 
will absent everywhere but the boundary of the plasma channel. Third, contrary to usual 
cases of laser filamentation where the medium is effectively transparent until the 
intensity is high enough to ionize the gas, here we have a resonantly absorbing medium until  
all the atoms have been completely ionized. At this point however, the medium is 
rendered almost transparent.  
All this means that we have a hybrid system - around the pulse edge, where intensity is small, we 
may 
expect phenomena familiar from resonant nonlinear optics 
\cite{Boshier1982,Lamb1971,Lamare1994,Delagnes2008}.
On the other hand, around the pulse center where intensity is large we may expect processes similar 
to the ones encountered in filamentation studies 
\cite{Fill1994,Couairon2007,Berge2007,Kolesik2013}. 

In order to investigate the propagation of 
ultra-short, ionizing laser pulses resonant with a transition from the atomic ground 
state in rubidium vapor, we develop a relatively simple model for the nonlinear optical response 
of the atoms and perform computer simulations to investigate propagation phenomena. 
We analyze the competing dynamics of self-focusing, nonlinear absorption and diffraction that 
govern the reshaping of the pulse in the medium and the geometry of the plasma channel left behind 
after the interaction. Our aim is to 
identify the requirements for the formation of a clean, continuous plasma channel 
with constant plasma density whose transverse dimensions are sufficient for use in plasma 
wake-field acceleration devices.

\section{Theory}

\subsection{Basic approach}

We set out to calculate the long range ($\sim$ 10 m) propagation of 780 nm wavelength, $\sim$ 100 
fs 
laser pulses in Rb vapor. The pulses are intense enough to ionize via multiphoton or tunnel 
ionization directly from the ground state ($I\sim \mathrm{~TW/cm^2})$, but are also 
resonant with the transition from the atomic ground state to the first excited state.   
The vapor density is $\sim 10^{14}-10^{15} \mathrm{~1/cm}^3$, far below the atmospheric densities 
usually considered in filamentation studies. In order to calculate pulse propagation,  
we need a wave equation for the light-field and couple it with 
the atomic response functions. The transient atomic response is expected to be dominated by the 
single photon resonances, so the traditional approach of using nonlinear 
susceptibility functions with various powers of the intensity does not work.
The classical formulas for anomalous dispersion in the vicinity of 
an absorption line are also useless at these timescales, we expect that Rabi-like oscillations will
yield the atomic response, augmented by ionization losses. Ab-initio methods 
that calculate the evolution of the electron wavefunction in space from a bound state to continuum 
states are theoretically sound and can treat this situation naturally, but are 
computationally too costly for using to calculate long range propagation 
and parameter scans. To make extended calculations feasible, we consider an axially symmetric 
system, physical quantities are assumed to depend only on the $r$ coordinate in the transverse 
plane.

\subsection{Model equations}

We assume that the laser field is linearly polarized and employ an envelope description of both the 
electric field $E$ and material response $P$ of the rubidium vapor, separating the central 
frequency of the laser: 
$E(\vec{r},z,t)=\frac{1}{2}\mathcal{E}(\vec{r},z,t)\exp(ik_0z-\omega_0t)+c.c.$ and  
$P(\vec{r},z,t)=\frac{1}{2}\mathcal{P}(\vec{r},z,t)\exp(ik_0z-\omega_0t)+c.c.$ . 
Here $z$ is the propagation direction, $\vec{r}$ is the position in the plane transverse to it 
and $\omega_0=k_0c$ the central frequency of the laser. The medium response is entirely contained in 
the polarization function $P(\vec{r},z,t)$, linear and nonlinear parts are not separated explicitly.
Using the standard transformation to a moving reference frame $\xi=z$, $\tau=t-z/c$, 
employing the paraxial approximation for propagation along $z$, rewriting the wave equation for the 
envelopes in frequency space 
$\tilde{\mathcal{E}}(\vec{r},\xi,\omega)=\mathfrak{F}\{\mathcal{E}(\vec{r},\xi,\tau)\}$,
$\tilde{\mathcal{P}}(\vec{r},\xi, \omega)=\mathfrak{F}\{\mathcal{P}(\vec{r},\xi,\tau)\}$ 
(where $\mathfrak{F}\{\ldotp\} $ denotes the time-Fourier transform) and finally 
employing the 
Slowly Evolving Wave Approximation (SEWA) \cite{Brabec1997,Couairon2011}  we arrive 
at the wave equation:
\begin{equation}
\begin{aligned}
\nabla_\perp^2\tilde{\mathcal{E}}(\vec{r},\xi,\omega) &
+i2(k_0+k)\partial_\xi\tilde{\mathcal{E}}(\vec{r},\xi,\omega)
= \\
&-(k_0+k)^2\tilde{\mathcal{P}}(\vec{r},\xi,\omega)/\epsilon_0. 
\end{aligned}
\label{waveeq}
\end{equation}
Here $k,\omega$ are the wavevector and the angular frequency of the various 
components offset 
from $k_0$ and $\omega_0$, $k=\omega/c$ and $\epsilon_0$ is the vacuum permittivity. 
The SEWA approximation that we use for deriving a first order wave equation has been developed for 
treating the propagation of ultra-short
(few-cycle) pulses and is much less restrictive than the Slowly Varying Envelope 
Approximation (SVEA) widely used in resonant nonlinear optics. In particular, it is still valid if 
the pulse develops a sharp leading edge during propagation.

\begin{figure}[htb]
\includegraphics[width=0.35\textwidth]{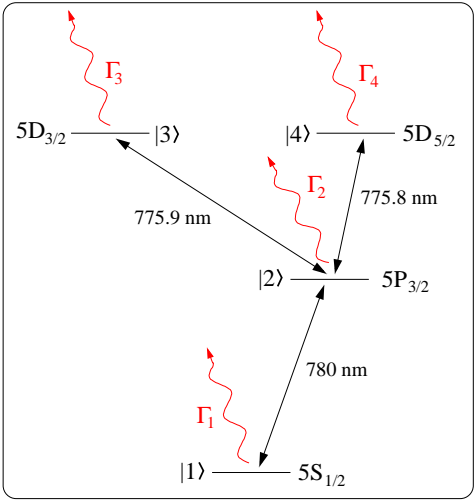}
\caption{Electronic levels of the Rb atom that are included in the model and their numbering. Three 
excited states are resonantly accessible from the ground state, ionization leads to level loss from 
each of the levels.}
\label{fig_Rblevels}       
\end{figure}

Atomic rubidium has a single valence electron outside a closed shell and an atomic transition from 
the $\mathrm{5S_{1/2}}$ ground state to the $\mathrm{5P_{3/2}}$ first excited state at 
$780\mathrm{~nm}$ (the $D_2$ line), precisely the same as the central wavelength of the 
Ti:sapphire laser used at the AWAKE experiment \cite{Gschwendtner2016}. Furthermore, there 
are two transitions from the first excited state to higher atomic states still well within 
the bandwidth of the laser: the $\mathrm{5P_{3/2}\rightarrow 5D_{3/2}}$ transition at 775.9 nm and 
the $\mathrm{5P_{3/2}\rightarrow 5D_{5/2}}$ one at 775.8 nm \cite{steckRb85,NIST}, see Fig. 
\ref{fig_Rblevels}. The transition from the ground state to $\mathrm{5P_{1/2}}$ at 794.8 nm is well 
out of resonance for this setup, so there are three excited states resonantly 
accessible from the ground state because of the coupling to the laser light.
A ``minimal'' model of the atom used to calculate the optical response that is 
lightweight enough to be employed in extended propagation calculations
must therefore include these four states as well as the process of photoionization 
from theses states.   

We start by separating the material response into parts describing atomic polarization 
due to resonant transitions between bound states and an absorption term due to ionization 
processes: $\mathcal{P}=\mathcal{P}_{atomic}+\mathcal{P}_{ionization}$.
We write the Schr{\"o}dinger equation for the probability amplitudes of the four quantum states 
in the spirit of resonant nonlinear optics. We take the quantization 
axis of the atomic angular momentum parallel  
to the direction of polarization, so the magnetic quantum number $m$ is conserved. Assuming the 
initial state of the atom to be in the ground state, without any 
constraint on generality we may use the set of atomic quantum states 
$\{|1\rangle=|5S_{1/2},m=1/2\rangle,
  |2\rangle=|5P_{3/2},m=1/2\rangle, 
|3\rangle=|5D_{3/2},m=1/2\rangle, |4\rangle=|5S_{5/2},m=1/2\rangle\}$ as an expansion basis for 
the atomic wave function with time dependent expansion coefficients $\alpha_j(t)$ 
(the other linkage pattern 
with $m=-1/2$ is symmetrical to this one). Note that this separation is possible because radiative 
transitions that could yield transitions 
between states of different $m$ are completely negligible on the $\sim$100 fs timescale. 
The wave function is thus written as: 
\begin{equation}
 |\psi(t)\rangle = \sum_{j=1}^4 \alpha_j(t) |j\rangle.
\end{equation}
Next we introduce the transformed probability amplitudes $a_j(t)$ by applying the phase 
transformation with respect to $\omega_0$ and $\omega_2$ as:
\begin{equation}
\begin{aligned}
a_1(t) & = \alpha_1(t) e^{i(\omega_2-\omega_0)t} \\
a_2(t) & = \alpha_2(t) e^{i\omega_2t} \\
a_3(t) & = \alpha_3(t) e^{i(\omega_2+\omega_0)t} \\
a_4(t) & = \alpha_4(t) e^{i(\omega_2+\omega_0)t} \\
\end{aligned}
\end{equation}
Here $\omega_2$ is the energy of the first excited state $|2\rangle$ divided by $\hbar$ and the 
transformation amounts choosing this energy as reference and to transforming to a reference fame 
rotating with the optical field. Using the Hamiltonian
\begin{equation}
 \hat{H}=\sum_{j=1}^4\hbar\omega_j|j\rangle\langle j| - \hat{d}E
\end{equation}
we obtain the equations for the probability amplitudes in the moving frame:
\begin{equation}
\begin{aligned}
 \partial_\tau a_1 =& -i\Delta_{21}a_1+\frac{i}{2}\Omega^*D_{21}a_2 - 
\frac{\Gamma_1}{2}a_1\\
 \partial_\tau a_2 =& \frac{i}{2}\bigl(\Omega D_{21}a_1 + 
   \Omega^*D_{32}a_3 + \Omega^*D_{42}a_4\bigr) - \frac{\Gamma_2}{2}a_2 \\
  \partial_\tau a_3 =& i\Delta_{32}a_3+\frac{i}{2}\Omega D_{32}a_2 - 
\frac{\Gamma_3}{2}a_3\\
 \partial_\tau a_4 =& i\Delta_{42}a_4+\frac{i}{2}\Omega D_{42}a_2 - \frac{\Gamma_4}{2}a_4
\end{aligned}
\label{schrodinger}
\end{equation}
Here we have introduced the notation $\Delta_{kl}=\omega_0-(\omega_k-\omega_l)$ for the detuning of 
the central laser frequency from the relevant atomic transitions and used the Rotating Wave 
approximation (RWA). We have also introduced the Rabi frequency for a unit dipole $\Omega(\tau)= 
\mathcal{E}(\tau)ea_0/\hbar$ ($e$ is the elementary charge and $a_0$ the Bohr radius) and written 
the dipole matrix elements in units of $ea_0$ as well, 
$\langle k|\hat{d}|l\rangle = D_{kl}ea_0$. $\Gamma_k$ 
are phenomenological loss terms for the 
level probabilities that describe photoionization and we have suppressed the explicit space and 
time 
dependence of $a_k$, $\Gamma_k$ and $\Omega$ for brevity. The material parameters $\omega_k$ and 
$D_{kl}$ are obtained from the literature \cite{NIST,steckRb85,Safronova2004}, 
their numerical values are quoted in the appendix. The intensity dependent photoionization rates 
$\Gamma_k$ for the two lower atomic levels $|1\rangle$ and $|2\rangle$are obtained from the 
so-called PPT formulas \cite{Perelomov1966,Perelomov1967,Perelomov1967b} that describe both 
multiphoton ionization and tunnel ionization in 
a unified way. For the two higher lying states $|3\rangle,|4\rangle$ an experimentally measured 
photoionization cross section is used as detailed in the appendix. Solving Eqs. \ref{schrodinger} 
at 
any point in space allows us to calculate the atomic part of the polarization 
$P_{atomic}=\langle \psi|\hat{d}|\psi\rangle$ for insertion into the wave equation.  

The wave equation in frequency space Eq. \ref{waveeq} is written in terms of 
$\tilde{\Omega}(\vec{r},\xi,\omega)$:
\begin{equation}
 \partial_\xi\tilde{\Omega} = 
\frac{i}{2}\frac{c}{\omega_0+\omega}\nabla_\perp^2\tilde{\Omega} + 
i\kappa_1\frac{\omega_0+\omega}{c}\tilde{p}-\kappa_2\tilde{\mathcal{Q}}.
\label{waveeq2}
\end{equation}
Here the first term describes diffraction, the second term is due to atomic polarization due to 
transitions between bound states: 
\begin{equation}
\tilde{p}=\mathfrak{F}\{p(\vec{r},\xi,\tau)\}= \mathfrak{F}\{D_{21}a_1^*a_2 + D_{23}a_2^*a_3
+ D_{24}a_2^*a_4\}.
\label{p}
\end{equation}
The third term corresponds to $P_{ionization}$ and is purely an energy loss term
derived from the requirement that the laser pulse should loose an appropriate number times the 
energy of a photon each time an atom is ionized:  
\begin{equation}
\tilde{\mathcal{Q}}=\mathfrak{F}\{\mathcal{Q}(\vec{r},\xi,\tau)\}=\mathfrak{F}\left\{\sum_j n_j 
\frac{\Gamma_j |a_j|^2}{\Omega^*}\right\}
\label{q}
\end{equation}
The numbers $n_j$ are the photon numbers associated with the ionization process from each of the 
atomic states. Note that they can be intensity dependent non-integers as the ionization rates may 
contain contributions from higher photon-number processes (see Eqs. \ref{ionrates_appendix}), 
though they are practically always close to the minimal number of required photons in our case.
The constants appearing in Eq. \ref{waveeq2} are given by:
\begin{equation}
 \begin{aligned}
\kappa_1=\frac{\mathcal{N}e^2a_0^2}{\hbar\epsilon_0} & , &
\kappa_2=\frac{\eta_0\omega_0\mathcal{N}e^2a_0^2}{\hbar} 
\end{aligned}
\end{equation}
where $\mathcal{N}$ is the vapor density and $\eta_0$ is the impedance of vacuum.
Equations \ref{schrodinger} and \ref{waveeq2} together with the relations \ref{p} and \ref{q} 
constitute the set of equations we have to solve for the investigation of our problem.

\section{Propagation calculations}

The equations were solved numerically assuming an axially symmetric 
system, i.e. all quantities 
were taken to depend on the propagation direction $z$ and the transverse radial coordinate $r$. 
The incident pulse was assumed to be a Gaussian beam with 
the waist located at $z=0$ the start of the interaction, the initial beam diameter $d$ 
(intensity FWHM width) and pulse energy $E_0$ being the two parameters varied during the parameter 
scans. The temporal shape of the incident pulse envelope was a hyperbolic secant 
$\mathrm{sech}(t/\tau_p)$ with $\tau_p = 85.0944\mathrm{~fs}$
which translates to a pulse duration of 150 fs. $\mathcal{N}=2\times 10^{14}/\mathrm{cm}^3$ vapor 
density was used in all calculations. Eqs. \ref{schrodinger} were solved with a fourth-order 
Dormand-Prince algorithm at each step of the numerical integration of Eq. \ref{waveeq2}. A 
split-step operator scheme was used for the latter equation.

\subsection{Pulse self-focusing and self-channeling}

\begin{figure}[htb]
\includegraphics[width=0.5\textwidth]{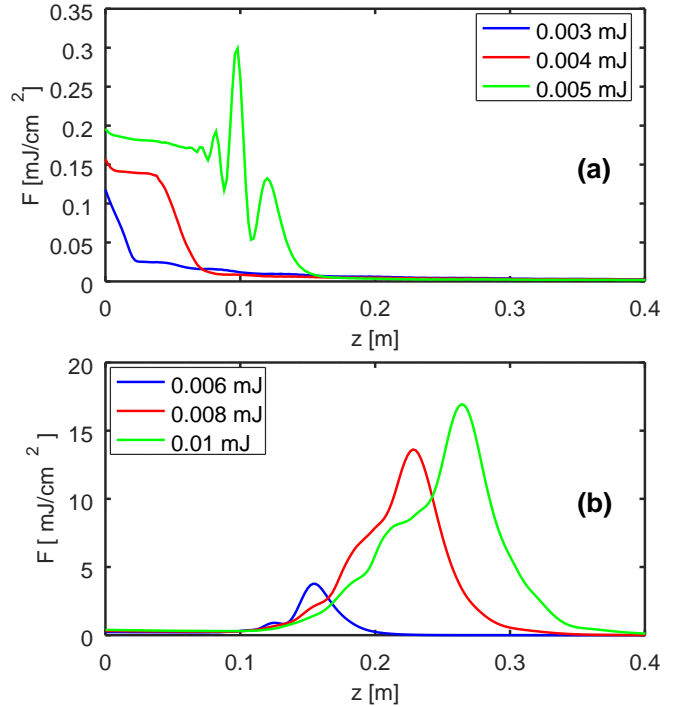}
\caption{On-axis radiant fluence $\mathcal{F}_0(z)$ in $\mathrm{mJ/cm^2}$ as a function of 
propagation distance $z$ for six values of the initial pulse energy $E_0$.}
\label{low_energy}       
\end{figure}

Pulse interaction with the rubidium vapor was first investigated for low energy pulses. With a beam 
waist diameter $d=1.5\mathrm{~mm}$ and pulse energies of 
$E_0=10^{-3}-10^{-2}\mathrm{~mJ}$, the 
initial on-axis peak intensity is $I\approx 10^9\mathrm{~W/cm^2}$, too small to ionize the 
atoms during the pulse. The spatial evolution of the on-axis radiant fluence 
$\mathcal{F}_0(z)=\int I(r=0,z,t)dt$ with propagation distance $z$ has been plotted 
for several values of $E_0$ in Fig. \ref{low_energy}. Traces of 
self focusing are visible even for $E_0=0.003 \mathrm{~mJ}$ as a marked deviation from 
an exponentially decreasing absorption curve (panel (a), blue curve) \textendash
 absorption clearly still dominates though. 
However, for $E_0=0.006 \mathrm{~mJ}$, we already have a pulse 
 focused around $z\approx 0.15\mathrm{~m}$ with the peak on-axis fluence over an order of 
magnitude greater than its initial ($z=0$) value, despite absorption (panel (b), blue curve). The 
overall behavior is very similar to that found for laser propagation in a medium of resonant
two-level atoms \cite{Boshier1982}, where the nonlinear refractive index and saturable absorption 
were both found to contribute to self-focusing. (Note however, that those results were derived 
for CW beams and ionization completely absent.) 

The onset of self-focusing here is 
considerably different from that caused by the classical intensity-dependent 
refractive index $n_2\cdot I$ in transparent 
media \cite{Kelley1965,Marburger1975}. First, the required pulse power is orders of magnitude 
smaller as the 0.006 mJ pulse plotted in panel (b) of Fig. \ref{low_energy} corresponds to 
$P=40\mathrm{~MW}$. Compared with the GW power required in atmospheric density gases 
\cite{Couairon2007} and noting that vapor density in our case is five orders of magnitude smaller, 
it is clear that the nonlinearity in this system is about $10^7$ times 
larger. Second, the location of the nonlinear focus increases with increasing pulse energy (or 
power) which is different from the scalings $(P/P_{cr})^{-1/2}$ and $(P/P_{cr})^{-1}$ observed 
for nonresonant pulses in various power domains \cite{Fibich2005}. Third, not only 
the overall pulse power, but also peak intensity and radiant fluence (and hence beam waist 
diameter) are important parameters in this system as the nonlinearity competes with both 
diffraction and 
absorption and it is easily saturated as atoms are lost from interaction via 
ionization. Indeed, for the $E_0=0.01\mathrm{~mJ}$ green curve in panel b) of Fig. 
\ref{low_energy}, ionization probability is already close to 80\% at the center of the nonlinear 
focus.   
 
\begin{figure}[htb]
\includegraphics[width=0.5\textwidth]{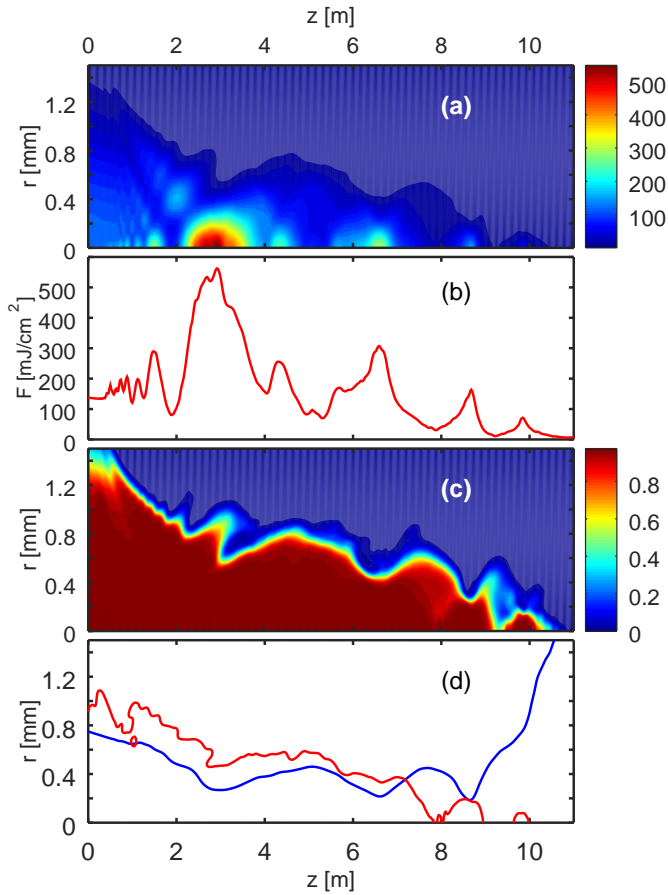}
\caption{The propagation of a laser pulse with $E_0=3.5\mathrm{~mJ}$ in Rb vapor. The horizontal 
axes of all four panels is 
$z$, the propagation distance, with identical scaling. (a) Contour plot of $\mathcal{F}(r,z)$ 
$[\mathrm{mJ/cm^2}]$. (b) On-axis fluence $\mathcal{F}_0(z)$. (c) Contour plot of the final 
ionization probability. (d) Blue line: half-energy radius $R_{1/2}(z)$ of the 
beam, red line: the boundary of 98\% ionization probability.}
\label{filament1}       
\end{figure}

Calculations for higher pulse energies yield interesting solutions that at first sight bear 
considerable 
resemblance to filamentation phenomena in air when self-focusing leads to plasma generation. 
A typical scenario is shown in Fig. \ref{filament1}. The spatial evolution of 
the radiant fluence is shown in panel (a), its on-axis value vs. propagation distance on panel (b). 
The plots clearly show that as the pulse propagates in the medium, the 
energy is focused periodically around the axis. The peaks decrease 
in amplitude and radial extension as the pulse progresses and energy is lost. 
The vapor is ionized completely close to the axis, the boundary of the plasma channel expanding 
and contracting repeatedly
with the radial extension of the laser pulse. (Panel c) displays the spatial 
distribution of the final ionization probability. Note that in this case the pulse is 
already intense enough to ionize the atoms at $z=0$, without self-focusing.) 
As the pulse energy is depleted, the plasma channel 
narrows and eventually ends as the pulse is no longer able to ionize the atoms. 
Clearly, there is a dynamic competition between nonlinear polarization, absorption and 
diffraction that yields an irregular, quasi-periodic plasma channel. 

A closer look reveals some fundamental differences to laser filamentation in air and other 
gases. In those scenarios the gas is essentially transparent, there is little or no loss when pulse 
intensity is not high enough to ionize the atoms or molecules. Self-focusing increases intensity 
until it is stopped (or rather dynamically balanced) by a combination of diffraction, plasma 
defocusing, strong energy losses due to multiphoton ionization, a saturation of $n_2$ or the 
emergence of higher order defocusing 
nonlinearities \cite{Feit1974,Couairon2003,Bejot2011,Couairon2007}. Because 
a large portion of the pulse energy can propagate outside the highly intense domain, the filament 
may regenerate even if its central, most intense portion is blocked 
\cite{Courvoisier2003,Kolesik2004,Skupin2004}. 
Conversely, in our case 
there is absorption for arbitrarily small intensities, but the absorber is easily saturable, the 
medium becomes transparent when it is fully ionized.  
Panel (d) of Fig.\ref{low_energy} displays two curves, the 
boundary of 98\% ionization (red line) which is a measure of the extent of the plasma channel
and the 'half-energy width' $R_{1/2}(z)$ of the laser beam (blue line). This latter is defined such 
that
\begin{equation} 
\int_0^{R_{1/2}} 2\pi r\mathcal{F}(r,z)dr=\frac{E(z)}{2}
\end{equation}
i.e. exactly half of the overall energy of the pulse at any given propagation distance $z$ is 
contained within the domain $r\leq R_{1/2}(z)$. (A beam width parameter like the FWHM in intensity 
or fluence would not be very representative as the 
beam cross-section does not remain a Gaussian and at certain positions it does not peak at 
$r=0$ but may have a hollow beam structure.) The figure shows that most of the pulse energy 
propagates within the plasma channel where absorption and nonlinear refraction are saturated. 
There is a self-channeling of the 
energy, self-focusing by the nonlinear medium is halted by the completion of the 
channel with full ionization where the laser field travels through a homogeneous, transparent 
plasma medium. There is no 
further absorption because the ionization potential of the second electron of rubidium is so much 
higher than that of the first one. Plasma defocusing within the channel is also absent as there is 
no gradient of plasma density within the channel core. Diffraction is the only mechanism that 
makes the beam expand repeatedly. Naturally, 
energy is constantly lost from the front part of the pulse as the plasma channel is created and 
eventually energy is depleted beyond a threshold that complete ionization ceases. This is marked by 
the crossing of the $R_{1/2}$ curve with the plasma channel boundary, the channel ends very close 
to the crossing. There may be one or two short ``revivals'' of plasma formation as remnants of the 
pulse refocus to ionize again, but compared to the length of the primary plasma channel, this 
distance is short, the propagation ends promptly after the plasma channel is interrupted for the 
first time. 

\begin{figure}[htb]
\includegraphics[width=0.5\textwidth]{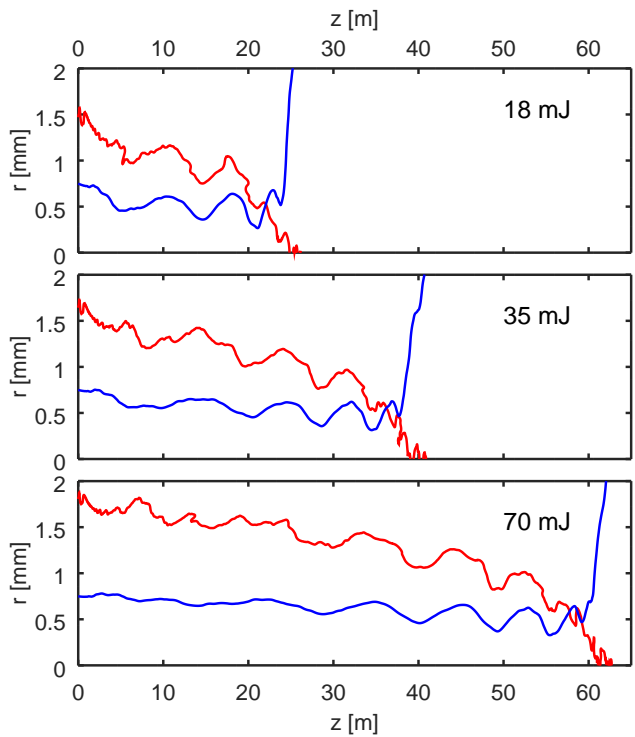}
\caption{Boundary of 98\% ionization (red lines) and $R_{1/2}(z)$ of the 
laser beam (blue lines) vs. propagation distance $z$ for three values of the incident pulse 
energy $E_0$. The horizontal axis has the same scale on all three panels.}
\label{high_energy}       
\end{figure}

\begin{figure}[htb]
\includegraphics[width=0.5\textwidth]{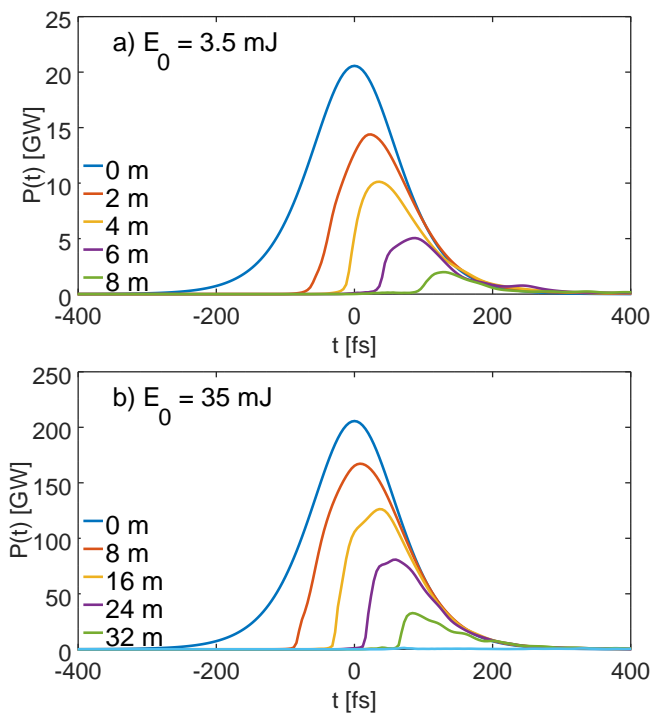}
\caption{Time evolution of the pulse power at several positions along the propagation for two 
values of the initial energy a) $E_0=3.5\mathrm{~mJ}$ and b)  $E_0=35\mathrm{~mJ}$.}
\label{pulse_reshape}       
\end{figure}

These general features are valid for pulses of higher energy as 
demonstrated by Fig.\ref{high_energy}, which depicts the same plot (ion channel 
radius and $R_{1/2}$ vs. propagation distance) for three different initial pulse energies $E_0$. 
The fact that the trailing part of the pulse propagates in the transparent plasma channel almost 
unchanged can be seen of Fig. \ref{pulse_reshape} where the temporal evolution of the pulse power 
(spatially integrated intensity) at several values of the propagation distance are plotted for two 
values of the initial pulse energy. The pulses are not attenuated homogeneously, energy 
is absorbed mostly around the leading edge (until full ionization is achieved). The leading edge 
steepens, while the trailing edge remains almost unchanged. 

\subsection{Plasma channel properties}

For the purposes of wakefield accelerator devices, the longitudinal and transverse extent of the 
plasma channel is of great importance, as is plasma homogeneity \textendash the channel must be 
continuous, sufficiently wide with very close to 100\% ionization. It can 
be seen on Fig. \ref{high_energy} that
while the channel radius fluctuates considerably as the 
pulse propagates, there is also a clear tendency of gradual narrowing until the pulse ``crashes'', 
i.e. the plasma channel radius becomes zero and the pulse intensity decreases below the level 
required for close to full ionization. Almost until this point the channel is uninterrupted, 
continuous and has a radius of $\sim$ 1 mm.
 
\begin{figure}[htb]
\includegraphics[width=0.5\textwidth]{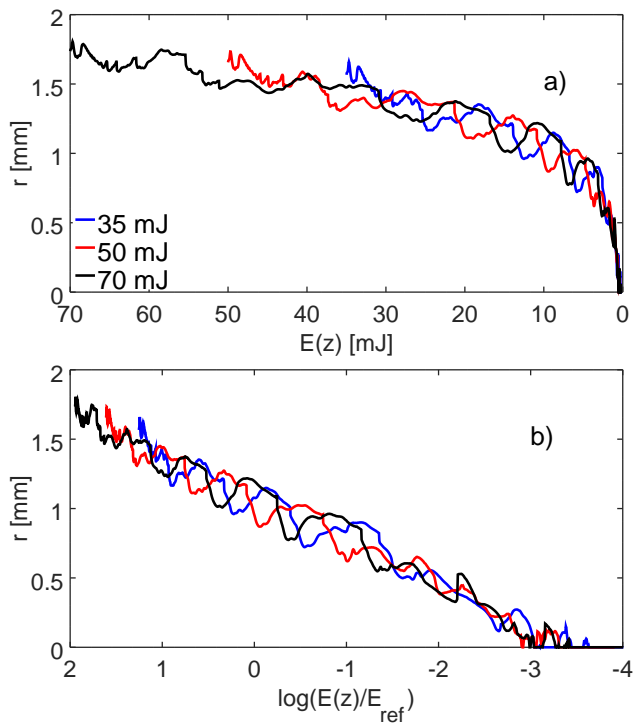}
\caption{a) Plasma channel radius (boundary of 98\%  ionization probability) as a function 
of the energy still contained in the pulse for three different values of initial pulse energy 
$E_0$, 35 mJ, 50 mJ and 70 mJ. b) Plasma channel radius as a function of $\log(E(z)/E_{ref})$. The 
reference energy is
$E_{ref}=10\mathrm{~mJ}$.}
\label{channel_radius}       
\end{figure}

To make a more quantitative comparison, the evolution of the plasma channel has been 
calculated for a large number of initial pulse energies and the channel radius 
(radius of 98\% ionization probability) plotted as a function of the energy $E(z)$ still left in 
the pulse after propagating a distance $z$. Some plots can be seen on Fig. \ref{channel_radius}, 
panel a). (The x axis of the plot has been reversed so that the pulses 
``propagate'' from left to right similar to the rest of the figures in the paper.) 
It is clear that the average radius of the plasma channel as well as the magnitude of the 
fluctuations around it are the same for pulses that possess the same energy 
during their propagation at their respective propagation distances. 
Only the ``phase'' of these quasi-periodic oscillations differ. The channels end rather abruptly 
close to $E(z)=0$ in a very similar manner in all three cases. 

The same quantity (98\% ionization probability radius) is plotted in Fig. 
\ref{channel_radius} panel b) as a function of $\log(E(z)/E_{ref})$ where we have taken 
$E_{ref}=10\mathrm{~mJ}$ as a reference. This shows that 
the average channel radius is linear in this quantity, all three 
curves oscillate around the same line to a very good approximation. In fact, a linear fit 
to the curves $r(x)=m x + r_0$ (where $x$ stands for 
$\log(E/E_{ref}$) yields very similar values: $m=0.298\pm0.005\mathrm{~mm}$ and 
$r_0=1.062\pm0.003\mathrm{~mm}$ when averaged over 9 calculations with $E_0$ values between 
$16\mathrm{~mJ}-70\mathrm{~mJ}$. This suggests that there is a global attractor to the behavior 
of the propagating pulse that is independent of the initial pulse energy in the domain 
investigated. Repeating the calculations with a different beam waist parameter ($d=2\mathrm{~mm}$ 
initial beam diameter) we obtain a similar behavior, but different parameters for the line of best 
fit for the $r$ vs. $\log(E(z)/E_{ref})$ curves, namely $m=0.350\pm0.007\mathrm{~mm}$. This 
indicates that though there is a globally attracting behavior also in this case, this is 
quantitatively different from the one for $d=1.5\mathrm{~mm}$, i.e. initial beam focusing has a 
long-term effect on the propagation.

\begin{figure}[htb]
\includegraphics[width=0.5\textwidth]{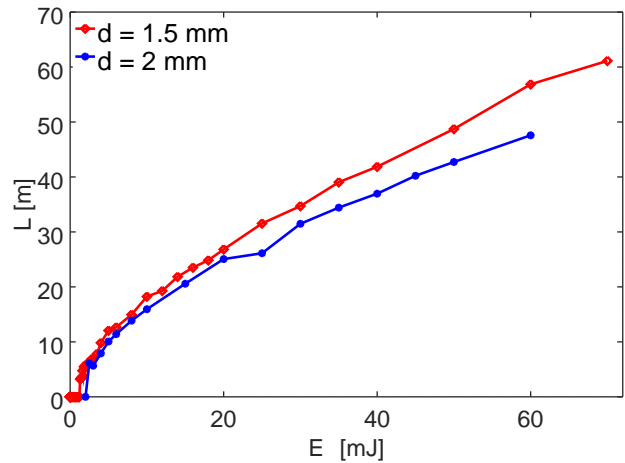}
\caption{Plasma channel length as a function of initial pulse energy $L(E_0)$  for two values of 
the initial beam diameter. }
\label{channel_length}       
\end{figure}

For our purposes, we will now define the length of the 
plasma channel $L$ as the propagation distance at which the on-axis ionization probability 
drops below 98\% for the first time. With this definition, the channel length is strictly 
zero for pulses that fail to ionize 98\% of the atoms at $z=0$, even if self-focusing increases 
on-axis intensity to create a channel after some propagation length. Clearly, $L$ will 
depend on the initial pulse energy and focusing (among other parameters) and, due to the nature of 
the radius curve with quasi-periodic oscillations, this quantity too will oscillate somewhat. 
Plotting $L$ as a function of the initial pulse energy for two different values of the initial laser 
beam diameter
 (Fig. \ref{channel_length}) shows that there is indeed a long-term effect of the initial focusing 
on the propagation. The difference between the two curves increases with $E_0$ which would not 
be the expected behavior if, after some initial transient the pulse  propagation tended to the 
same attractor solution for both beam diameters.

\subsection{Effects of initial focusing}

\begin{figure}[htb]
\includegraphics[width=0.5\textwidth]{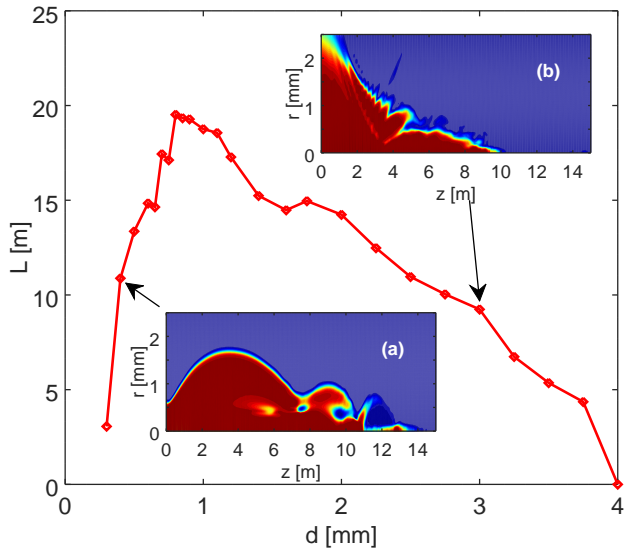}
\caption{Main plot: Plasma channel length as a function of initial beam diameter $L(d)$ for a $E_0 
= 8\mathrm{~mJ}$ pulse. Inset (a): contour plot of ionization probability as a function of radial 
distance $r$ and propagation distance $z$ for $d=0.4\mathrm{~mm}$.  Inset (b): contour plot of 
ionization probability for $d=3\mathrm{~mm}$.  }
\label{initial_focusing}       
\end{figure}

To investigate the effect of initial beam focusing, a set of calculations with constant $E_0$ 
but different $d$ was performed. Figure \ref{initial_focusing} depicts a curve of the plasma channel 
length for $E_0=8\mathrm{~mJ}$ pulses as a function of  $d$ (red line). The 
curve is not strictly monotonous, because the plasma channel length as defined above may change 
abruptly when, for certain parameters there is a small dip in the on-axis ionization probability 
close to the end of the pulse propagation before a revival of the ionization probability. However 
there is a clear maximum at $d=0.8\mathrm{~mm}$ and the channel length is a fraction of the 
maximum value when $d$ is much smaller or much larger than optimal. Two insets on Fig. 
\ref{initial_focusing} depict a contour plot of the ionization probability for two  
sub-optimal values of the initial beam diameter and reveal the reason for this behavior. 
When the initial focusing is too tight (inset (a) ), the Rayleigh range is small and diffraction 
causes the beam to expand and ionize in a larger radius around the axis,  
depleting the energy severely. When the 
initial spot size is too large on the other hand (inset (b) ), the initial channel radius is large 
and a lot of energy is lost before the beam contracts to a more modest size. 
An additional feature visible on the plots are the ``holes'' in the ionization profile, small 
localized domains where ionization is not perfect, plasma density is inhomogeneous within the 
channel. Therefore a good choice of initial focusing also proves to be important for realizing 
homogeneous, long plasma channels for plasma wave acceleration.

\section{Some further comments}

The model for the optical response of the atoms presented in this paper contains numerous 
approximations, 
trying to capture resonant interaction and ionization simultaneously and, at the same time, to 
be lightweight enough for extended propagation calculations in two spatial dimensions. In 
experiments and calculations of 
laser pulse filamentation in atmospheric gases it was observed that for sufficiently large values 
of the pulse power (several times the critical power $P_{cr}$ required for the onset 
of self-focusing and filamentation), a transverse instability breaks axial symmetry,
 the beam breaks up into multiple filaments 
\cite{Bespalov1966,Kandidov2005,Alonso2010,Champeaux2005,Rohwetter2008,Champeaux2008}. Our 
axially symmetric description naturally excludes obtaining such solutions. While not immediately 
obvious whether multifilamentation can appear in a resonant system, this means that our model may 
overestimate the plasma channel length for a given vapor density and laser focusing. 

The fact that the laser can resonantly transfer atoms to excited states has an effect on the 
ionization process as well. Analysis of the numerical solution shows that the onset of ionization 
is less abrupt, it starts at lower intensities than without resonance. 
The reason is that apart from three-photon ionization from the ground state, a process of 
resonant excitation followed by two-photon ionization, and a process of resonant excitation twice 
followed by single-photon ionization is also possible. In fact during the initial part of the 
propagation, before the pulse leading edge steepens too much, the dominant route to ionization is 
the one by two-photon absorption from the first excited state.

Initial derivation of the theory included an electron current term $\sim\partial J/\partial t$
that describes plasma absorption and dispersion. However, after verifying that this term has a very 
little effect in calculations presented in the paper, the term was neglected while performing  
extended parameter scans. This might be surprising at first, because in general, plasma density 
gradients are a major source of defocusing processes in laser pulse filamentation. 
However, in our case: {\em i)} because of full ionization 
the existence of plasma density gradients is limited to a narrow boundary region around the channel 
core, (in the center the plasma is completely homogeneous,) and most of the energy carried by 
the pulse is channeled in the transparent central part. {\em ii)} the vapor 
density is so low that even with full ionization, plasma is orders of magnitude less dense than in 
normal filamentation scenarios. Thus the fact that this term should have a negligible effect
on the shape and extension of the plasma channel is understandable. Under different conditions,
(e.g. much higher vapor densities or possibly much longer propagation lengths) the effects of the 
plasma term would be non-negligible.

Interaction of the laser pulse with the ionic core of the singly ionized rubidium has also been 
neglected completely in the present description. This is justified by the fact that the vapor is 
too rare for the usual, nonresonant optical coefficients to be effective and that the ionization 
potential is an order of magnitude greater than that for the valence electron of rubidium. 

The phenomena discussed in this paper should not be termed ``filamentation'' as it is 
understood in the usual sense. Filamentation in that sense occurs when there is an (almost) lossless 
Kerr medium to self-focus the beam and an abrupt onset of absorption due to multiphoton ionization.
An ionization potential much 
larger than the photon energy ($E_i\gg\hbar\omega$) is required for this \cite{Couairon2007} and 
the beam collapses to a transverse size $\sim 100 {\mathrm\mu m}$. In the present case absorption 
is always present except when the medium is saturated and the transverse size of the beam remains 
about an order of magnitude larger as the focusing nonlinearity is also saturable. This width is 
already sufficient for application in accelerator devices such as AWAKE.

\section{Summary}

We have investigated the propagation of ultra-short, ionizing laser pulses in rubidium vapor under 
conditions of direct single-photon resonance with an atomic transition from the ground state. To 
make the problem tractable for numerical solution in two spatial dimensions, we developed a 
relatively lightweight theory that includes the nonlinear response of atoms to resonant fields to 
all orders. Ionization was inserted in the theory as a phenomenological probability loss from the 
atomic levels. A split-step quasi-spectral method was used to solve a first order propagation 
equation in frequency space in the slowly-evolving wave approximation. 

The dynamics of self-focusing, plasma channel formation and pulse collapse due to energy depletion 
were studied using parameter scans of computer simulations. We have shown that given sufficient 
pulse energy, a competition between nonlinear self-focusing and diffraction results in the pulse 
energy being confined in a narrow region around the propagation axis. The front part of the 
propagating pulse ionizes atoms close to the axis and so a plasma channel is formed with almost 
complete ionization of the rubidium vapor. 
The energy of the trailing 
part of the pulse is guided along the channel which is essentially 
transparent for the field. The radius of the plasma channel exhibits quasi-periodic oscillations 
around an average value which in turn is determined by the energy remaining in the pulse at the 
given propagation distance. Initial pulse focusing has a long-term effect on propagation, the 
average channel radius is different for pulses with different initial beam widths even at large 
propagation distances. The dependence of the plasma channel length on initial pulse energy and beam 
diameter has been studied.  The calculations are expected to be useful for considerations in 
wakefield accelerator devices where the creation of homogeneous, spatially extended, dense plasmas 
are necessary, such as at the AWAKE project at CERN.

\begin{acknowledgments}
We thank Joshua T. Moody, Gagik P. Djotyan, Andrea Armaroli and J\'{e}r\^{o}me Kasparian for 
helpful discussions. 
This work was supported by the Excellence program EXMET of the Hungarian
Academy of Sciences under grant 2018-1.2.1-NKP-2018-00012 (National Excellence Program).
The use of the MTA Cloud (https://cloud.mta.hu/) facility was indispensible for the numerical 
computations and its use through the {\em Awakelaser} project is gratefully acknowledged.
Data processing a display was performed with the use of GNU Octave \cite{Octave}.

\end{acknowledgments}

\appendix

\section{Material parameters of the theory}

We use the following notation to identify atomic states in the equations:
\begin{equation*}
 \begin{aligned}
|5S_{1/2},m=1/2\rangle&\rightarrow |1\rangle\\
|5P_{3/2},m=1/2\rangle&\rightarrow |2\rangle\\
|5D_{3/2},m=1/2\rangle&\rightarrow |3\rangle\\
|5D_{5/2},m=1/2\rangle&\rightarrow |4\rangle  
 \end{aligned}
\end{equation*}
The energy levels of the excited states relative to the $|1\rangle$ ground state are \cite{NIST}:
$\hbar\omega_2 = 1.589049\mathrm{~eV}$, $\hbar\omega_3 = 3.1864603\mathrm{~eV}$,
$\hbar\omega_4 = 3.1868276\mathrm{~eV}$. 
Given that the photon energy for 780 nm light is 1.5895 eV, three photons are required for 
ionization from $|1\rangle$, two photons from $|2\rangle$ and a single photon from $|3\rangle$ and 
$|4\rangle$.
The dipole matrix elements between the states are obtained from \cite{steckRb85} and 
\cite{Safronova2004} and transformed to the conventions used in \cite{steckbook}:
 \begin{equation*}
 \begin{aligned}
\langle 1|\hat{d}|2\rangle =& 5.9786\cdot\mathrm{e\cdot a_0}\cdot\sqrt{1/4}\sqrt{2/3}\\
\langle 2|\hat{d}|3\rangle =& 0.787\cdot\mathrm{e\cdot 
a_0}\cdot\sqrt{1/4}\cdot\sqrt{1/15}\\
\langle 2|\hat{d}|4\rangle =& 2.334\cdot\mathrm{e\cdot 
a_0}\cdot\sqrt{1/6}\cdot\sqrt{3/5}
\end{aligned}
\end{equation*}
For the (intensity dependent) multiphoton ionization rates $\Gamma_1,\Gamma_2$ from the ground and 
first excited states we use the well known PPT formulas 
\cite{Perelomov1966,Perelomov1967,Perelomov1967b}. They are written using the notations of
\cite{Bejot2008}, and are reproduced below from Eqs. 1.25-1.30 on 
pages 19-21 of \cite{Bejot2008} for reference. The full formula for the ionization rate of any atom 
from a quantum state characterized by $l$ and $m_l$ is:
\begin{equation}
\begin{aligned}
  W(\omega_0,\gamma) = & \omega_{a.u.}\sqrt{\frac{6}{\pi}}|C_{n^*,l^*}|^2 f(l,m_l)
\frac{U_i}{2U_H}A_{m_l}(\omega_0,\gamma) \\
& \times\left(\frac{2E_0}{E\sqrt{1+\gamma^2}}\right)^{2n-|m_l|-3/2}
\mathrm{exp}\left[-\frac{2E_0}{3E}g(\gamma)\right]\\
\label{ionrate}
 \end{aligned}
\end{equation}
In this formula $\gamma$ is the famous Keldysh parameter
\begin{equation}
 \omega_0\frac{\sqrt{2m_eU_i}}{|eE_{max}|}
\end{equation}
with $U_i$ being the ionization energy, $U_i=4.177128\mathrm{~eV}$ for the $\mathrm{5S_{1/2}}$ 
state and $U_i=2.588079\mathrm{~eV}$ for the $\mathrm{5P_{3/2}}$ state. $m_e$ is the electron mass 
and $E_{max}$ is the maximum field amplitude. In \ref{ionrate} $U_H$ is the ionization energy of 
hydrogen, $\omega_{a.u.}=eE_H/\sqrt{2m_eU_H}\simeq4.1\cdot10^{16}\mathrm{~s^{-1}}$, 
$E_H=e^5m_e^2/(64\hbar^4\pi^3\epsilon_0^3)\simeq501.4\mathrm{~GV/m}$, $E_0=E_H(U_i/U_H)^{3/2}$.
The factor
\begin{equation}
|C_{n^*,l^*}|^2 = \frac{2^{2n^*}}{n^*\Gamma(n^*+l^*+1)\Gamma(n^*-l^*)}
\end{equation}
contains the effective quantum numbers $n^*$ which is $n^*=\sqrt{U_H/U_i}$ for $Z=1$, and 
$l^*=n^*-1$, $\Gamma()$ is the gamma function here. The rest of the factors in \ref{ionrate} are: 
\begin{equation}
\begin{aligned}
f(l,m_l) = & \frac{(2l+1)(l+|m_l|)!}{2^{|m_l|}(|m_l|)!(l-|m_l|)!} \\
A_{m_l}(\omega_0,\gamma) = & \frac{4\gamma^2}{\sqrt{3\pi}|m_l|!(1+\gamma^2)} \\
 & \times \sum_{K\geq\nu}^\infty e^{-\alpha(K-\nu)} \Phi_{m_l}\left(\sqrt{\beta(K-\nu)}\right)\\
\Phi_{m_l}(x) = & e^{-x^2}\int_0^x(x^2-y^2)^{|m_l|}e^{y^2}dy \\
\beta(\gamma) = & \frac{2\gamma}{\sqrt{1+\gamma^2}} \\
\alpha(\gamma) = & 2\mathrm{sinh}^{-1}(\gamma)-\beta(\gamma) \\
g(\gamma) = & \frac{3}{2\gamma}\left[ \left(1+\frac{1}{2\gamma^2}\right) 
\mathrm{sinh}^{-1}(\gamma)-\frac{\sqrt{1+\gamma^2}}{2\gamma}\right]\\
\nu_0 = & \frac{U_i}{\hbar\omega_0} \\
 \nu = & \nu_0\left(1+\frac{1}{2\gamma^2}\right)
\label{ionrates_appendix}
\end{aligned}
\end{equation}
The ionization rates calculated from Eq. \ref{ionrate} are used as $\Gamma_1$ and $\Gamma_2$ in 
Eqs. \ref{schrodinger} and \ref{q}. 
The single photon ionization rate $\Gamma_4$ was 
calculated using the experimental cross-section $\sigma=10.9 \mathrm{~Mb}$ from \cite{Duncan2001}, 
which has been measured for $\lambda=788 \mathrm{~nm}$ light. The same value was used for 
$\Gamma_3$.

\clearpage

\bibliography{/home/gdemeter/fiz/manuscript/bibliography_pulseprop}

\begin{thebibliography}{42}%
\makeatletter
\providecommand \@ifxundefined [1]{%
 \@ifx{#1\undefined}
}%
\providecommand \@ifnum [1]{%
 \ifnum #1\expandafter \@firstoftwo
 \else \expandafter \@secondoftwo
 \fi
}%
\providecommand \@ifx [1]{%
 \ifx #1\expandafter \@firstoftwo
 \else \expandafter \@secondoftwo
 \fi
}%
\providecommand \natexlab [1]{#1}%
\providecommand \enquote  [1]{``#1''}%
\providecommand \bibnamefont  [1]{#1}%
\providecommand \bibfnamefont [1]{#1}%
\providecommand \citenamefont [1]{#1}%
\providecommand \href@noop [0]{\@secondoftwo}%
\providecommand \href [0]{\begingroup \@sanitize@url \@href}%
\providecommand \@href[1]{\@@startlink{#1}\@@href}%
\providecommand \@@href[1]{\endgroup#1\@@endlink}%
\providecommand \@sanitize@url [0]{\catcode `\\12\catcode `\$12\catcode
  `\&12\catcode `\#12\catcode `\^12\catcode `\_12\catcode `\%12\relax}%
\providecommand \@@startlink[1]{}%
\providecommand \@@endlink[0]{}%
\providecommand \url  [0]{\begingroup\@sanitize@url \@url }%
\providecommand \@url [1]{\endgroup\@href {#1}{\urlprefix }}%
\providecommand \urlprefix  [0]{URL }%
\providecommand \Eprint [0]{\href }%
\providecommand \doibase [0]{http://dx.doi.org/}%
\providecommand \selectlanguage [0]{\@gobble}%
\providecommand \bibinfo  [0]{\@secondoftwo}%
\providecommand \bibfield  [0]{\@secondoftwo}%
\providecommand \translation [1]{[#1]}%
\providecommand \BibitemOpen [0]{}%
\providecommand \bibitemStop [0]{}%
\providecommand \bibitemNoStop [0]{.\EOS\space}%
\providecommand \EOS [0]{\spacefactor3000\relax}%
\providecommand \BibitemShut  [1]{\csname bibitem#1\endcsname}%
\let\auto@bib@innerbib\@empty
\bibitem [{\citenamefont {Bergé}(1998)}]{Berge1998}%
  \BibitemOpen
  \bibfield  {author} {\bibinfo {author} {\bibfnamefont {L.}~\bibnamefont
  {Bergé}},\ }\href {\doibase https://doi.org/10.1016/S0370-1573(97)00092-6}
  {\bibfield  {journal} {\bibinfo  {journal} {Physics Reports}\ }\textbf
  {\bibinfo {volume} {303}},\ \bibinfo {pages} {259 } (\bibinfo {year}
  {1998})}\BibitemShut {NoStop}%
\bibitem [{\citenamefont {Couairon}\ and\ \citenamefont
  {Mysyrowicz}(2007)}]{Couairon2007}%
  \BibitemOpen
  \bibfield  {author} {\bibinfo {author} {\bibfnamefont {A.}~\bibnamefont
  {Couairon}}\ and\ \bibinfo {author} {\bibfnamefont {A.}~\bibnamefont
  {Mysyrowicz}},\ }\href {\doibase
  https://doi.org/10.1016/j.physrep.2006.12.005} {\bibfield  {journal}
  {\bibinfo  {journal} {Physics Reports}\ }\textbf {\bibinfo {volume} {441}},\
  \bibinfo {pages} {47 } (\bibinfo {year} {2007})}\BibitemShut {NoStop}%
\bibitem [{\citenamefont {Berg{\'{e}}}\ \emph {et~al.}(2007)\citenamefont
  {Berg{\'{e}}}, \citenamefont {Skupin}, \citenamefont {Nuter}, \citenamefont
  {Kasparian},\ and\ \citenamefont {Wolf}}]{Berge2007}%
  \BibitemOpen
  \bibfield  {author} {\bibinfo {author} {\bibfnamefont {L.}~\bibnamefont
  {Berg{\'{e}}}}, \bibinfo {author} {\bibfnamefont {S.}~\bibnamefont {Skupin}},
  \bibinfo {author} {\bibfnamefont {R.}~\bibnamefont {Nuter}}, \bibinfo
  {author} {\bibfnamefont {J.}~\bibnamefont {Kasparian}}, \ and\ \bibinfo
  {author} {\bibfnamefont {J.-P.}\ \bibnamefont {Wolf}},\ }\href {\doibase
  10.1088/0034-4885/70/10/r03} {\bibfield  {journal} {\bibinfo  {journal}
  {Reports on Progress in Physics}\ }\textbf {\bibinfo {volume} {70}},\
  \bibinfo {pages} {1633} (\bibinfo {year} {2007})}\BibitemShut {NoStop}%
\bibitem [{\citenamefont {Kandidov}\ \emph {et~al.}(2009)\citenamefont
  {Kandidov}, \citenamefont {Shlenov},\ and\ \citenamefont
  {Kosareva}}]{Kandidov2009}%
  \BibitemOpen
  \bibfield  {author} {\bibinfo {author} {\bibfnamefont {V.~P.}\ \bibnamefont
  {Kandidov}}, \bibinfo {author} {\bibfnamefont {S.~A.}\ \bibnamefont
  {Shlenov}}, \ and\ \bibinfo {author} {\bibfnamefont {O.~G.}\ \bibnamefont
  {Kosareva}},\ }\href {\doibase 10.1070/QE2009v039n03ABEH013916} {\bibfield
  {journal} {\bibinfo  {journal} {Quantum Electronics}\ }\textbf {\bibinfo
  {volume} {39}},\ \bibinfo {pages} {205} (\bibinfo {year} {2009})}\BibitemShut
  {NoStop}%
\bibitem [{\citenamefont {Kolesik}\ and\ \citenamefont
  {Moloney}(2013)}]{Kolesik2013}%
  \BibitemOpen
  \bibfield  {author} {\bibinfo {author} {\bibfnamefont {M.}~\bibnamefont
  {Kolesik}}\ and\ \bibinfo {author} {\bibfnamefont {J.~V.}\ \bibnamefont
  {Moloney}},\ }\href {\doibase 10.1088/0034-4885/77/1/016401} {\bibfield
  {journal} {\bibinfo  {journal} {Reports on Progress in Physics}\ }\textbf
  {\bibinfo {volume} {77}},\ \bibinfo {pages} {016401} (\bibinfo {year}
  {2013})}\BibitemShut {NoStop}%
\bibitem [{\citenamefont {Caldwell}\ \emph {et~al.}(2016)\citenamefont
  {Caldwell} \emph {et~al.}}]{Caldwell2016}%
  \BibitemOpen
  \bibfield  {author} {\bibinfo {author} {\bibfnamefont {A.}~\bibnamefont
  {Caldwell}} \emph {et~al.},\ }\href {\doibase
  https://doi.org/10.1016/j.nima.2015.12.050} {\bibfield  {journal} {\bibinfo
  {journal} {Nuclear Instruments and Methods \allowbreak in Physics Research
  Section A: Accelerators, \allowbreak Spectrometers, \allowbreak Detectors and
  Associated Equipment}\ }\textbf {\bibinfo {volume} {829}},\ \bibinfo {pages}
  {3 } (\bibinfo {year} {2016})},\ \bibinfo {note} {2nd European Advanced
  Accelerator Concepts Workshop - EAAC 2015}\BibitemShut {NoStop}%
\bibitem [{\citenamefont {Gschwendtner}\ \emph {et~al.}(2016)\citenamefont
  {Gschwendtner} \emph {et~al.}}]{Gschwendtner2016}%
  \BibitemOpen
  \bibfield  {author} {\bibinfo {author} {\bibfnamefont {E.}~\bibnamefont
  {Gschwendtner}} \emph {et~al.},\ }\href {\doibase
  https://doi.org/10.1016/j.nima.2016.02.026} {\bibfield  {journal} {\bibinfo
  {journal} {Nuclear Instruments and Methods \allowbreak in Physics Research
  Section A: Accelerators, \allowbreak Spectrometers, \allowbreak Detectors and
  Associated Equipment}\ }\textbf {\bibinfo {volume} {829}},\ \bibinfo {pages}
  {76 } (\bibinfo {year} {2016})},\ \bibinfo {note} {2nd European Advanced
  Accelerator Concepts Workshop - EAAC 2015}\BibitemShut {NoStop}%
\bibitem [{\citenamefont {Adli}\ \emph {et~al.}(2018)\citenamefont {Adli},
  \citenamefont {Ahuja}, \citenamefont {Apsimon}, \citenamefont {Apsimon},
  \citenamefont {Bachmann}, \citenamefont {Barrientos}, \citenamefont {Batsch},
  \citenamefont {Bauche}, \citenamefont {Olsen}, \citenamefont {Bernardini}
  \emph {et~al.}}]{Adli2018}%
  \BibitemOpen
  \bibfield  {author} {\bibinfo {author} {\bibfnamefont {E.}~\bibnamefont
  {Adli}}, \bibinfo {author} {\bibfnamefont {A.}~\bibnamefont {Ahuja}},
  \bibinfo {author} {\bibfnamefont {O.}~\bibnamefont {Apsimon}}, \bibinfo
  {author} {\bibfnamefont {R.}~\bibnamefont {Apsimon}}, \bibinfo {author}
  {\bibfnamefont {A.-M.}\ \bibnamefont {Bachmann}}, \bibinfo {author}
  {\bibfnamefont {D.}~\bibnamefont {Barrientos}}, \bibinfo {author}
  {\bibfnamefont {F.}~\bibnamefont {Batsch}}, \bibinfo {author} {\bibfnamefont
  {J.}~\bibnamefont {Bauche}}, \bibinfo {author} {\bibfnamefont {V.~B.}\
  \bibnamefont {Olsen}}, \bibinfo {author} {\bibfnamefont {M.}~\bibnamefont
  {Bernardini}},  \emph {et~al.},\ }\href@noop {} {\bibfield  {journal}
  {\bibinfo  {journal} {Nature}\ }\textbf {\bibinfo {volume} {561}},\ \bibinfo
  {pages} {363} (\bibinfo {year} {2018})}\BibitemShut {NoStop}%
\bibitem [{\citenamefont {Doussot}\ \emph {et~al.}(2016)\citenamefont
  {Doussot}, \citenamefont {B\'ejot},\ and\ \citenamefont
  {Faucher}}]{Doussot2016}%
  \BibitemOpen
  \bibfield  {author} {\bibinfo {author} {\bibfnamefont {J.}~\bibnamefont
  {Doussot}}, \bibinfo {author} {\bibfnamefont {P.}~\bibnamefont {B\'ejot}}, \
  and\ \bibinfo {author} {\bibfnamefont {O.}~\bibnamefont {Faucher}},\ }\href
  {\doibase 10.1103/PhysRevA.94.013805} {\bibfield  {journal} {\bibinfo
  {journal} {Phys. Rev. A}\ }\textbf {\bibinfo {volume} {94}},\ \bibinfo
  {pages} {013805} (\bibinfo {year} {2016})}\BibitemShut {NoStop}%
\bibitem [{\citenamefont {Doussot}\ \emph {et~al.}(2017)\citenamefont
  {Doussot}, \citenamefont {Karras}, \citenamefont {Billard}, \citenamefont
  {B\'{e}jot},\ and\ \citenamefont {Faucher}}]{Doussot2017}%
  \BibitemOpen
  \bibfield  {author} {\bibinfo {author} {\bibfnamefont {J.}~\bibnamefont
  {Doussot}}, \bibinfo {author} {\bibfnamefont {G.}~\bibnamefont {Karras}},
  \bibinfo {author} {\bibfnamefont {F.}~\bibnamefont {Billard}}, \bibinfo
  {author} {\bibfnamefont {P.}~\bibnamefont {B\'{e}jot}}, \ and\ \bibinfo
  {author} {\bibfnamefont {O.}~\bibnamefont {Faucher}},\ }\href {\doibase
  10.1364/OPTICA.4.000764} {\bibfield  {journal} {\bibinfo  {journal} {Optica}\
  }\textbf {\bibinfo {volume} {4}},\ \bibinfo {pages} {764} (\bibinfo {year}
  {2017})}\BibitemShut {NoStop}%
\bibitem [{\citenamefont {{Boshier}}\ and\ \citenamefont
  {{Sandle}}(1982)}]{Boshier1982}%
  \BibitemOpen
  \bibfield  {author} {\bibinfo {author} {\bibfnamefont {M.~G.}\ \bibnamefont
  {{Boshier}}}\ and\ \bibinfo {author} {\bibfnamefont {W.~J.}\ \bibnamefont
  {{Sandle}}},\ }\href {\doibase 10.1016/0030-4018(82)90251-6} {\bibfield
  {journal} {\bibinfo  {journal} {Optics Communications}\ }\textbf {\bibinfo
  {volume} {42}},\ \bibinfo {pages} {371} (\bibinfo {year} {1982})}\BibitemShut
  {NoStop}%
\bibitem [{\citenamefont {LAMB}(1971)}]{Lamb1971}%
  \BibitemOpen
  \bibfield  {author} {\bibinfo {author} {\bibfnamefont {G.~L.}\ \bibnamefont
  {LAMB}},\ }\href {\doibase 10.1103/RevModPhys.43.99} {\bibfield  {journal}
  {\bibinfo  {journal} {Rev. Mod. Phys.}\ }\textbf {\bibinfo {volume} {43}},\
  \bibinfo {pages} {99} (\bibinfo {year} {1971})}\BibitemShut {NoStop}%
\bibitem [{\citenamefont {de~Lamare}\ \emph {et~al.}(1994)\citenamefont
  {de~Lamare}, \citenamefont {Comte},\ and\ \citenamefont
  {Kupecek}}]{Lamare1994}%
  \BibitemOpen
  \bibfield  {author} {\bibinfo {author} {\bibfnamefont {J.}~\bibnamefont
  {de~Lamare}}, \bibinfo {author} {\bibfnamefont {M.}~\bibnamefont {Comte}}, \
  and\ \bibinfo {author} {\bibfnamefont {P.}~\bibnamefont {Kupecek}},\ }\href
  {\doibase 10.1103/PhysRevA.50.3366} {\bibfield  {journal} {\bibinfo
  {journal} {Phys. Rev. A}\ }\textbf {\bibinfo {volume} {50}},\ \bibinfo
  {pages} {3366} (\bibinfo {year} {1994})}\BibitemShut {NoStop}%
\bibitem [{\citenamefont {Delagnes}\ and\ \citenamefont
  {Bouchene}(2008)}]{Delagnes2008}%
  \BibitemOpen
  \bibfield  {author} {\bibinfo {author} {\bibfnamefont {J.}~\bibnamefont
  {Delagnes}}\ and\ \bibinfo {author} {\bibfnamefont {M.}~\bibnamefont
  {Bouchene}},\ }\href {\doibase https://doi.org/10.1016/j.optcom.2008.08.047}
  {\bibfield  {journal} {\bibinfo  {journal} {Optics Communications}\ }\textbf
  {\bibinfo {volume} {281}},\ \bibinfo {pages} {5824 } (\bibinfo {year}
  {2008})}\BibitemShut {NoStop}%
\bibitem [{\citenamefont {Fill}(1994)}]{Fill1994}%
  \BibitemOpen
  \bibfield  {author} {\bibinfo {author} {\bibfnamefont {E.~E.}\ \bibnamefont
  {Fill}},\ }\href {\doibase 10.1364/JOSAB.11.002241} {\bibfield  {journal}
  {\bibinfo  {journal} {J. Opt. Soc. Am. B}\ }\textbf {\bibinfo {volume}
  {11}},\ \bibinfo {pages} {2241} (\bibinfo {year} {1994})}\BibitemShut
  {NoStop}%
\bibitem [{\citenamefont {Brabec}\ and\ \citenamefont
  {Krausz}(1997)}]{Brabec1997}%
  \BibitemOpen
  \bibfield  {author} {\bibinfo {author} {\bibfnamefont {T.}~\bibnamefont
  {Brabec}}\ and\ \bibinfo {author} {\bibfnamefont {F.}~\bibnamefont
  {Krausz}},\ }\href {\doibase 10.1103/PhysRevLett.78.3282} {\bibfield
  {journal} {\bibinfo  {journal} {Phys. Rev. Lett.}\ }\textbf {\bibinfo
  {volume} {78}},\ \bibinfo {pages} {3282} (\bibinfo {year}
  {1997})}\BibitemShut {NoStop}%
\bibitem [{\citenamefont {Couairon}\ \emph {et~al.}(2011)\citenamefont
  {Couairon}, \citenamefont {Brambilla}, \citenamefont {Corti}, \citenamefont
  {Majus}, \citenamefont {de~J.~Ram{\'i}rez-G{\'o}ngora},\ and\ \citenamefont
  {Kolesik}}]{Couairon2011}%
  \BibitemOpen
  \bibfield  {author} {\bibinfo {author} {\bibfnamefont {A.}~\bibnamefont
  {Couairon}}, \bibinfo {author} {\bibfnamefont {E.}~\bibnamefont {Brambilla}},
  \bibinfo {author} {\bibfnamefont {T.}~\bibnamefont {Corti}}, \bibinfo
  {author} {\bibfnamefont {D.}~\bibnamefont {Majus}}, \bibinfo {author}
  {\bibfnamefont {O.}~\bibnamefont {de~J.~Ram{\'i}rez-G{\'o}ngora}}, \ and\
  \bibinfo {author} {\bibfnamefont {M.}~\bibnamefont {Kolesik}},\ }\href
  {\doibase 10.1140/epjst/e2011-01503-3} {\bibfield  {journal} {\bibinfo
  {journal} {The European Physical Journal Special Topics}\ }\textbf {\bibinfo
  {volume} {199}},\ \bibinfo {pages} {5} (\bibinfo {year} {2011})}\BibitemShut
  {NoStop}%
\bibitem [{\citenamefont {Steck}(2009)}]{steckRb85}%
  \BibitemOpen
  \bibfield  {author} {\bibinfo {author} {\bibfnamefont {D.~A.}\ \bibnamefont
  {Steck}},\ }\href {http://steck.us/alkalidata} {\emph {\bibinfo {title}
  {Rubidium 85 D Line Data}}}\ (\bibinfo  {publisher} {available online at
  http://steck.us/alkalidata},\ \bibinfo {year} {(revision 2.1.2, 12 August
  2009)})\BibitemShut {NoStop}%
\bibitem [{\citenamefont {Kramida}\ \emph {et~al.}(2018)\citenamefont
  {Kramida}, \citenamefont {Ralchenko}, \citenamefont {Reader},\ and\
  \citenamefont {Team}}]{NIST}%
  \BibitemOpen
  \bibfield  {author} {\bibinfo {author} {\bibfnamefont {A.}~\bibnamefont
  {Kramida}}, \bibinfo {author} {\bibfnamefont {Y.}~\bibnamefont {Ralchenko}},
  \bibinfo {author} {\bibfnamefont {J.}~\bibnamefont {Reader}}, \ and\ \bibinfo
  {author} {\bibfnamefont {N.~A.}\ \bibnamefont {Team}},\ }\href
  {https://physics.nist.gov/asd} {\emph {\bibinfo {title} {NIST Atomic Spectra
  Database}}}\ (\bibinfo  {publisher} {National Institute of Standards and
  Technology, Gaithersburg, MD.},\ \bibinfo {year} {2018})\BibitemShut
  {NoStop}%
\bibitem [{\citenamefont {Safronova}\ \emph {et~al.}(2004)\citenamefont
  {Safronova}, \citenamefont {Williams},\ and\ \citenamefont
  {Clark}}]{Safronova2004}%
  \BibitemOpen
  \bibfield  {author} {\bibinfo {author} {\bibfnamefont {M.~S.}\ \bibnamefont
  {Safronova}}, \bibinfo {author} {\bibfnamefont {C.~J.}\ \bibnamefont
  {Williams}}, \ and\ \bibinfo {author} {\bibfnamefont {C.~W.}\ \bibnamefont
  {Clark}},\ }\href {\doibase 10.1103/PhysRevA.69.022509} {\bibfield  {journal}
  {\bibinfo  {journal} {Phys. Rev. A}\ }\textbf {\bibinfo {volume} {69}},\
  \bibinfo {pages} {022509} (\bibinfo {year} {2004})}\BibitemShut {NoStop}%
\bibitem [{\citenamefont {Perelomov}\ \emph {et~al.}(1966)\citenamefont
  {Perelomov}, \citenamefont {Popov},\ and\ \citenamefont
  {Terent'ev}}]{Perelomov1966}%
  \BibitemOpen
  \bibfield  {author} {\bibinfo {author} {\bibfnamefont {A.~M.}\ \bibnamefont
  {Perelomov}}, \bibinfo {author} {\bibfnamefont {V.~S.}\ \bibnamefont
  {Popov}}, \ and\ \bibinfo {author} {\bibfnamefont {M.~V.}\ \bibnamefont
  {Terent'ev}},\ }\href@noop {} {\bibfield  {journal} {\bibinfo  {journal}
  {Soviet Physics JETP}\ }\textbf {\bibinfo {volume} {23}},\ \bibinfo {pages}
  {924} (\bibinfo {year} {1966})}\BibitemShut {NoStop}%
\bibitem [{\citenamefont {Perelomov}\ \emph {et~al.}(1967)\citenamefont
  {Perelomov}, \citenamefont {Popov},\ and\ \citenamefont
  {Terent'ev}}]{Perelomov1967}%
  \BibitemOpen
  \bibfield  {author} {\bibinfo {author} {\bibfnamefont {A.~M.}\ \bibnamefont
  {Perelomov}}, \bibinfo {author} {\bibfnamefont {V.~S.}\ \bibnamefont
  {Popov}}, \ and\ \bibinfo {author} {\bibfnamefont {M.~V.}\ \bibnamefont
  {Terent'ev}},\ }\href@noop {} {\bibfield  {journal} {\bibinfo  {journal}
  {Soviet Physics JETP}\ }\textbf {\bibinfo {volume} {24}},\ \bibinfo {pages}
  {207} (\bibinfo {year} {1967})}\BibitemShut {NoStop}%
\bibitem [{\citenamefont {Perelomov}\ and\ \citenamefont
  {Popov}(1967)}]{Perelomov1967b}%
  \BibitemOpen
  \bibfield  {author} {\bibinfo {author} {\bibfnamefont {A.~M.}\ \bibnamefont
  {Perelomov}}\ and\ \bibinfo {author} {\bibfnamefont {V.~S.}\ \bibnamefont
  {Popov}},\ }\href@noop {} {\bibfield  {journal} {\bibinfo  {journal} {Soviet
  Physics JETP}\ }\textbf {\bibinfo {volume} {25}},\ \bibinfo {pages} {336}
  (\bibinfo {year} {1967})}\BibitemShut {NoStop}%
\bibitem [{\citenamefont {Kelley}(1965)}]{Kelley1965}%
  \BibitemOpen
  \bibfield  {author} {\bibinfo {author} {\bibfnamefont {P.~L.}\ \bibnamefont
  {Kelley}},\ }\href {\doibase 10.1103/PhysRevLett.15.1005} {\bibfield
  {journal} {\bibinfo  {journal} {Phys. Rev. Lett.}\ }\textbf {\bibinfo
  {volume} {15}},\ \bibinfo {pages} {1005} (\bibinfo {year}
  {1965})}\BibitemShut {NoStop}%
\bibitem [{\citenamefont {{Marburger}}(1975)}]{Marburger1975}%
  \BibitemOpen
  \bibfield  {author} {\bibinfo {author} {\bibfnamefont {J.~H.}\ \bibnamefont
  {{Marburger}}},\ }\href {\doibase 10.1016/0079-6727(75)90003-8} {\bibfield
  {journal} {\bibinfo  {journal} {Progress in Quantum Electronics}\ }\textbf
  {\bibinfo {volume} {4}},\ \bibinfo {pages} {35} (\bibinfo {year}
  {1975})}\BibitemShut {NoStop}%
\bibitem [{\citenamefont {Fibich}\ \emph {et~al.}(2005)\citenamefont {Fibich},
  \citenamefont {Eisenmann}, \citenamefont {Ilan}, \citenamefont {Erlich},
  \citenamefont {Fraenkel}, \citenamefont {Henis}, \citenamefont {Gaeta},\ and\
  \citenamefont {Zigler}}]{Fibich2005}%
  \BibitemOpen
  \bibfield  {author} {\bibinfo {author} {\bibfnamefont {G.}~\bibnamefont
  {Fibich}}, \bibinfo {author} {\bibfnamefont {S.}~\bibnamefont {Eisenmann}},
  \bibinfo {author} {\bibfnamefont {B.}~\bibnamefont {Ilan}}, \bibinfo {author}
  {\bibfnamefont {Y.}~\bibnamefont {Erlich}}, \bibinfo {author} {\bibfnamefont
  {M.}~\bibnamefont {Fraenkel}}, \bibinfo {author} {\bibfnamefont
  {Z.}~\bibnamefont {Henis}}, \bibinfo {author} {\bibfnamefont {A.~L.}\
  \bibnamefont {Gaeta}}, \ and\ \bibinfo {author} {\bibfnamefont
  {A.}~\bibnamefont {Zigler}},\ }\href {\doibase 10.1364/OPEX.13.005897}
  {\bibfield  {journal} {\bibinfo  {journal} {Opt. Express}\ }\textbf {\bibinfo
  {volume} {13}},\ \bibinfo {pages} {5897} (\bibinfo {year}
  {2005})}\BibitemShut {NoStop}%
\bibitem [{\citenamefont {Feit}\ and\ \citenamefont {Fleck}(1974)}]{Feit1974}%
  \BibitemOpen
  \bibfield  {author} {\bibinfo {author} {\bibfnamefont {M.~D.}\ \bibnamefont
  {Feit}}\ and\ \bibinfo {author} {\bibfnamefont {J.~A.}\ \bibnamefont
  {Fleck}},\ }\href {\doibase 10.1063/1.1655139} {\bibfield  {journal}
  {\bibinfo  {journal} {Applied Physics Letters}\ }\textbf {\bibinfo {volume}
  {24}},\ \bibinfo {pages} {169} (\bibinfo {year} {1974})}\BibitemShut
  {NoStop}%
\bibitem [{\citenamefont {Couairon}(2003)}]{Couairon2003}%
  \BibitemOpen
  \bibfield  {author} {\bibinfo {author} {\bibfnamefont {A.}~\bibnamefont
  {Couairon}},\ }\href {\doibase 10.1103/PhysRevA.68.015801} {\bibfield
  {journal} {\bibinfo  {journal} {Phys. Rev. A}\ }\textbf {\bibinfo {volume}
  {68}},\ \bibinfo {pages} {015801} (\bibinfo {year} {2003})}\BibitemShut
  {NoStop}%
\bibitem [{\citenamefont {B\'ejot}\ \emph {et~al.}(2011)\citenamefont
  {B\'ejot}, \citenamefont {Hertz}, \citenamefont {Kasparian}, \citenamefont
  {Lavorel}, \citenamefont {Wolf},\ and\ \citenamefont {Faucher}}]{Bejot2011}%
  \BibitemOpen
  \bibfield  {author} {\bibinfo {author} {\bibfnamefont {P.}~\bibnamefont
  {B\'ejot}}, \bibinfo {author} {\bibfnamefont {E.}~\bibnamefont {Hertz}},
  \bibinfo {author} {\bibfnamefont {J.}~\bibnamefont {Kasparian}}, \bibinfo
  {author} {\bibfnamefont {B.}~\bibnamefont {Lavorel}}, \bibinfo {author}
  {\bibfnamefont {J.~P.}\ \bibnamefont {Wolf}}, \ and\ \bibinfo {author}
  {\bibfnamefont {O.}~\bibnamefont {Faucher}},\ }\href {\doibase
  10.1103/PhysRevLett.106.243902} {\bibfield  {journal} {\bibinfo  {journal}
  {Phys. Rev. Lett.}\ }\textbf {\bibinfo {volume} {106}},\ \bibinfo {pages}
  {243902} (\bibinfo {year} {2011})}\BibitemShut {NoStop}%
\bibitem [{\citenamefont {Courvoisier}\ \emph {et~al.}(2003)\citenamefont
  {Courvoisier}, \citenamefont {Boutou}, \citenamefont {Kasparian},
  \citenamefont {Salmon}, \citenamefont {Méjean}, \citenamefont {Yu},\ and\
  \citenamefont {Wolf}}]{Courvoisier2003}%
  \BibitemOpen
  \bibfield  {author} {\bibinfo {author} {\bibfnamefont {F.}~\bibnamefont
  {Courvoisier}}, \bibinfo {author} {\bibfnamefont {V.}~\bibnamefont {Boutou}},
  \bibinfo {author} {\bibfnamefont {J.}~\bibnamefont {Kasparian}}, \bibinfo
  {author} {\bibfnamefont {E.}~\bibnamefont {Salmon}}, \bibinfo {author}
  {\bibfnamefont {G.}~\bibnamefont {Méjean}}, \bibinfo {author} {\bibfnamefont
  {J.}~\bibnamefont {Yu}}, \ and\ \bibinfo {author} {\bibfnamefont {J.-P.}\
  \bibnamefont {Wolf}},\ }\href {\doibase 10.1063/1.1592615} {\bibfield
  {journal} {\bibinfo  {journal} {Applied Physics Letters}\ }\textbf {\bibinfo
  {volume} {83}},\ \bibinfo {pages} {213} (\bibinfo {year} {2003})}\BibitemShut
  {NoStop}%
\bibitem [{\citenamefont {Kolesik}\ and\ \citenamefont
  {Moloney}(2004)}]{Kolesik2004}%
  \BibitemOpen
  \bibfield  {author} {\bibinfo {author} {\bibfnamefont {M.}~\bibnamefont
  {Kolesik}}\ and\ \bibinfo {author} {\bibfnamefont {J.~V.}\ \bibnamefont
  {Moloney}},\ }\href {\doibase 10.1364/OL.29.000590} {\bibfield  {journal}
  {\bibinfo  {journal} {Opt. Lett.}\ }\textbf {\bibinfo {volume} {29}},\
  \bibinfo {pages} {590} (\bibinfo {year} {2004})}\BibitemShut {NoStop}%
\bibitem [{\citenamefont {Skupin}\ \emph {et~al.}(2004)\citenamefont {Skupin},
  \citenamefont {Berg\'e}, \citenamefont {Peschel},\ and\ \citenamefont
  {Lederer}}]{Skupin2004}%
  \BibitemOpen
  \bibfield  {author} {\bibinfo {author} {\bibfnamefont {S.}~\bibnamefont
  {Skupin}}, \bibinfo {author} {\bibfnamefont {L.}~\bibnamefont {Berg\'e}},
  \bibinfo {author} {\bibfnamefont {U.}~\bibnamefont {Peschel}}, \ and\
  \bibinfo {author} {\bibfnamefont {F.}~\bibnamefont {Lederer}},\ }\href
  {\doibase 10.1103/PhysRevLett.93.023901} {\bibfield  {journal} {\bibinfo
  {journal} {Phys. Rev. Lett.}\ }\textbf {\bibinfo {volume} {93}},\ \bibinfo
  {pages} {023901} (\bibinfo {year} {2004})}\BibitemShut {NoStop}%
\bibitem [{\citenamefont {{Bespalov}}\ and\ \citenamefont
  {{Talanov}}(1966)}]{Bespalov1966}%
  \BibitemOpen
  \bibfield  {author} {\bibinfo {author} {\bibfnamefont {V.~I.}\ \bibnamefont
  {{Bespalov}}}\ and\ \bibinfo {author} {\bibfnamefont {V.~I.}\ \bibnamefont
  {{Talanov}}},\ }\href@noop {} {\bibfield  {journal} {\bibinfo  {journal}
  {Soviet Journal of Experimental and Theoretical Physics Letters}\ }\textbf
  {\bibinfo {volume} {3}},\ \bibinfo {pages} {307} (\bibinfo {year}
  {1966})}\BibitemShut {NoStop}%
\bibitem [{\citenamefont {Kandidov}\ \emph {et~al.}(2005)\citenamefont
  {Kandidov}, \citenamefont {Akozbek}, \citenamefont {Scalora}, \citenamefont
  {Kosareva}, \citenamefont {Nyakk}, \citenamefont {Luo}, \citenamefont
  {Hosseini},\ and\ \citenamefont {Chin}}]{Kandidov2005}%
  \BibitemOpen
  \bibfield  {author} {\bibinfo {author} {\bibfnamefont {V.}~\bibnamefont
  {Kandidov}}, \bibinfo {author} {\bibfnamefont {N.}~\bibnamefont {Akozbek}},
  \bibinfo {author} {\bibfnamefont {M.}~\bibnamefont {Scalora}}, \bibinfo
  {author} {\bibfnamefont {O.}~\bibnamefont {Kosareva}}, \bibinfo {author}
  {\bibfnamefont {A.}~\bibnamefont {Nyakk}}, \bibinfo {author} {\bibfnamefont
  {Q.}~\bibnamefont {Luo}}, \bibinfo {author} {\bibfnamefont {S.}~\bibnamefont
  {Hosseini}}, \ and\ \bibinfo {author} {\bibfnamefont {S.}~\bibnamefont
  {Chin}},\ }\href {\doibase 10.1007/s00340-004-1677-1} {\bibfield  {journal}
  {\bibinfo  {journal} {Applied Physics B}\ }\textbf {\bibinfo {volume} {80}},\
  \bibinfo {pages} {267} (\bibinfo {year} {2005})}\BibitemShut {NoStop}%
\bibitem [{\citenamefont {Alonso}\ \emph {et~al.}(2010)\citenamefont {Alonso},
  \citenamefont {Za\"{i}r}, \citenamefont {Rom\'{a}n}, \citenamefont {Varela},\
  and\ \citenamefont {Roso}}]{Alonso2010}%
  \BibitemOpen
  \bibfield  {author} {\bibinfo {author} {\bibfnamefont {B.}~\bibnamefont
  {Alonso}}, \bibinfo {author} {\bibfnamefont {A.}~\bibnamefont {Za\"{i}r}},
  \bibinfo {author} {\bibfnamefont {J.~S.}\ \bibnamefont {Rom\'{a}n}}, \bibinfo
  {author} {\bibfnamefont {O.}~\bibnamefont {Varela}}, \ and\ \bibinfo {author}
  {\bibfnamefont {L.}~\bibnamefont {Roso}},\ }\href {\doibase
  10.1364/OE.18.015467} {\bibfield  {journal} {\bibinfo  {journal} {Opt.
  Express}\ }\textbf {\bibinfo {volume} {18}},\ \bibinfo {pages} {15467}
  (\bibinfo {year} {2010})}\BibitemShut {NoStop}%
\bibitem [{\citenamefont {Champeaux}\ and\ \citenamefont
  {Berg\'e}(2005)}]{Champeaux2005}%
  \BibitemOpen
  \bibfield  {author} {\bibinfo {author} {\bibfnamefont {S.}~\bibnamefont
  {Champeaux}}\ and\ \bibinfo {author} {\bibfnamefont {L.}~\bibnamefont
  {Berg\'e}},\ }\href {\doibase 10.1103/PhysRevE.71.046604} {\bibfield
  {journal} {\bibinfo  {journal} {Phys. Rev. E}\ }\textbf {\bibinfo {volume}
  {71}},\ \bibinfo {pages} {046604} (\bibinfo {year} {2005})}\BibitemShut
  {NoStop}%
\bibitem [{\citenamefont {Rohwetter}\ \emph {et~al.}(2008)\citenamefont
  {Rohwetter}, \citenamefont {Quei\ss{}er}, \citenamefont {Stelmaszczyk},
  \citenamefont {Fechner},\ and\ \citenamefont {W\"oste}}]{Rohwetter2008}%
  \BibitemOpen
  \bibfield  {author} {\bibinfo {author} {\bibfnamefont {P.}~\bibnamefont
  {Rohwetter}}, \bibinfo {author} {\bibfnamefont {M.}~\bibnamefont
  {Quei\ss{}er}}, \bibinfo {author} {\bibfnamefont {K.}~\bibnamefont
  {Stelmaszczyk}}, \bibinfo {author} {\bibfnamefont {M.}~\bibnamefont
  {Fechner}}, \ and\ \bibinfo {author} {\bibfnamefont {L.}~\bibnamefont
  {W\"oste}},\ }\href {\doibase 10.1103/PhysRevA.77.013812} {\bibfield
  {journal} {\bibinfo  {journal} {Phys. Rev. A}\ }\textbf {\bibinfo {volume}
  {77}},\ \bibinfo {pages} {013812} (\bibinfo {year} {2008})}\BibitemShut
  {NoStop}%
\bibitem [{\citenamefont {Champeaux}\ \emph {et~al.}(2008)\citenamefont
  {Champeaux}, \citenamefont {Berg\'e}, \citenamefont {Gordon}, \citenamefont
  {Ting}, \citenamefont {Pe\~nano},\ and\ \citenamefont
  {Sprangle}}]{Champeaux2008}%
  \BibitemOpen
  \bibfield  {author} {\bibinfo {author} {\bibfnamefont {S.}~\bibnamefont
  {Champeaux}}, \bibinfo {author} {\bibfnamefont {L.}~\bibnamefont {Berg\'e}},
  \bibinfo {author} {\bibfnamefont {D.}~\bibnamefont {Gordon}}, \bibinfo
  {author} {\bibfnamefont {A.}~\bibnamefont {Ting}}, \bibinfo {author}
  {\bibfnamefont {J.}~\bibnamefont {Pe\~nano}}, \ and\ \bibinfo {author}
  {\bibfnamefont {P.}~\bibnamefont {Sprangle}},\ }\href {\doibase
  10.1103/PhysRevE.77.036406} {\bibfield  {journal} {\bibinfo  {journal} {Phys.
  Rev. E}\ }\textbf {\bibinfo {volume} {77}},\ \bibinfo {pages} {036406}
  (\bibinfo {year} {2008})}\BibitemShut {NoStop}%
\bibitem [{\citenamefont {Eaton}\ \emph {et~al.}(2017)\citenamefont {Eaton},
  \citenamefont {Bateman}, \citenamefont {Hauberg},\ and\ \citenamefont
  {Wehbring}}]{Octave}%
  \BibitemOpen
  \bibfield  {author} {\bibinfo {author} {\bibfnamefont {J.~W.}\ \bibnamefont
  {Eaton}}, \bibinfo {author} {\bibfnamefont {D.}~\bibnamefont {Bateman}},
  \bibinfo {author} {\bibfnamefont {S.}~\bibnamefont {Hauberg}}, \ and\
  \bibinfo {author} {\bibfnamefont {R.}~\bibnamefont {Wehbring}},\ }\href
  {https://www.gnu.org/software/octave/doc/v4.2.1/} {\emph {\bibinfo {title}
  {{GNU Octave} version 4.2.1 manual: a high-level interactive language for
  numerical computations}}} (\bibinfo {year} {2017})\BibitemShut {NoStop}%
\bibitem [{\citenamefont {Steck}(2016)}]{steckbook}%
  \BibitemOpen
  \bibfield  {author} {\bibinfo {author} {\bibfnamefont {D.~A.}\ \bibnamefont
  {Steck}},\ }\href {http://steck.us/teaching} {\emph {\bibinfo {title}
  {Quantum and Atom Optics}}}\ (\bibinfo  {publisher} {available online at
  http://steck.us/teaching},\ \bibinfo {year} {(revision 0.10.3, 13 April
  2016)})\BibitemShut {NoStop}%
\bibitem [{\citenamefont {B\'{e}jot}(2008)}]{Bejot2008}%
  \BibitemOpen
  \bibfield  {author} {\bibinfo {author} {\bibfnamefont {P.}~\bibnamefont
  {B\'{e}jot}},\ }\emph {\bibinfo {title} {Theoretical and experimental
  investigations of ultrashort laser filamentation in gases}},\ \href {\doibase
  DOI: 10.13097/archive-ouverte/unige:4568} {Ph.D. thesis} (\bibinfo {year}
  {2008}),\ \bibinfo {note} {, Universit\'{e} de Gen\`{e}ve 2008
  https://archive-ouverte.unige.ch/unige:4568}\BibitemShut {NoStop}%
\bibitem [{\citenamefont {Duncan}\ \emph {et~al.}(2001)\citenamefont {Duncan},
  \citenamefont {Sanchez-Villicana}, \citenamefont {Gould},\ and\ \citenamefont
  {Sadeghpour}}]{Duncan2001}%
  \BibitemOpen
  \bibfield  {author} {\bibinfo {author} {\bibfnamefont {B.~C.}\ \bibnamefont
  {Duncan}}, \bibinfo {author} {\bibfnamefont {V.}~\bibnamefont
  {Sanchez-Villicana}}, \bibinfo {author} {\bibfnamefont {P.~L.}\ \bibnamefont
  {Gould}}, \ and\ \bibinfo {author} {\bibfnamefont {H.~R.}\ \bibnamefont
  {Sadeghpour}},\ }\href {\doibase 10.1103/PhysRevA.63.043411} {\bibfield
  {journal} {\bibinfo  {journal} {Phys. Rev. A}\ }\textbf {\bibinfo {volume}
  {63}},\ \bibinfo {pages} {043411} (\bibinfo {year} {2001})}\BibitemShut
  {NoStop}%
\end{thebibliography}%

\bibliographystyle{apsrev4-1}


\end{document}